# Occupational Diversity and Stratification in Platform Work: A Longitudinal Study of Online Freelancers (Pre-Print)

PYEONGHWA KIM, Syracuse University, USA
TAYLOR LEWANDOWSKI, Syracuse University, USA
MICHAEL DUNN, Skidmore College, USA
STEVE SAWYER, Syracuse University, USA

We focus on occupational diversity in platform-mediated work to advance conceptual and empirical insight into the occupationally embedded nature of platform labor. We pursue this focus in response to a prevailing tendency to treat platform workers as a homogeneous group, overlooking the unique demands, constraints, and practices rooted in specific professions. Such generalizations hinder both understanding of platform work and the development of sociotechnical systems that support differentiated occupational realities. To address this gap, we present a longitudinal analysis of 108 online freelancers spanning five occupational categories. We show that occupational context structures workers' capacity to interpret and navigate platformic management, shaping distinct experiences across four dimensions of platform work: self-presentation, flexibility, skilling, and platform work sustainability. To articulate how digital labor platforms' managerial control interacts with occupational embeddedness, we introduce the concept of platformic occupational stratification and discuss four mechanisms that explain its logic and implications for platform-mediated work. These insights contribute to CSCW by informing occupation-sensitive research and design approaches that directly engage with the specific opportunities and challenges rooted in workers' situated occupational agency in platform-mediated work.

## 1 INTRODUCTION

We present a longitudinal analysis examining *how workers' experiences vary by occupational contexts in platform-mediated work*. We do so because of the growth in both the types and volume of platform-mediated work being done [23,104]. We situate our study in sociological accounts that treat occupations as loci of social power and stratification [39,108,114]. This lens helps reveal occupationally differentiated experiences and underlying sociotechnical mechanisms within platform-mediated labor, which are often obscured when computer-supported cooperative work (CSCW) and human–computer interaction (HCI) research amalgamates diverse occupational distinctions into a single notion of platform workers [56,101]. In doing so, we contribute to CSCW and HCI scholarship by advancing a more nuanced conceptual and analytical lens for understanding and designing for occupationally embedded user experiences in platform-mediated work.

Like others, we define an occupation as a work arrangement in which workers typically perform tasks that require cognate skills and are typically labeled by synonymous job titles [5]. At least eight occupational categories span more than 500 digital labor platforms, accounting for about 12.5% of the global workforce [23]. Despite conceptual recognition of occupational heterogeneity in platform-mediated labor, existing CSCW and HCI literature often approaches platform workers as a uniform group (e.g., gig workers, platform workers, online freelancers), thereby limiting the advancement of knowledge on how occupational contexts shape differential use cases, challenges, and design requirements [53,61,62,101]. This tendency leads to a growing scholarly call for research approaches that are sensitive to the varied worker experiences across occupational lines [42,55,101,109,112].

Specifically, we are motivated by two intersecting discourses within CSCW, HCI, and occupational sociology: (1) the rise of diversity-sensitive research and (2) the conceptual and analytical value of occupation as a structural context shaping workers' platform labor experiences. The diversity-sensitive research movement embodies the concept of social situatedness of workers within the third wave of HCI research, which views diversity not as an external attribute of 'workers' but as a constitutive property of the interactional setting where user interfaces, interaction flows, and worker identities co-emerge in use [15,36,38]. Diversity-sensitive research in CSCW and HCI calls for systematic attention to how people's social identities and roles contour their interactions with computing systems [42]. Ethnographies of work in the CSCW and HCI communities also provide an empirical basis for the emergence of such movements by examining the embedding of worker-technology interactions in particular empirical sites such as London Underground line control rooms [40] and hospital operation wards [8]. These studies show that workers use shared technological artifacts in occupation-specific ways, with distinct responsibilities, specialized tasks, and situational demands structuring technology-in-use and its design implications. In the same vein, recent platform work research investigates specific occupations, such as food deliverers [89,90,115], beauticians [6,7], wellness professionals [84], care workers [99], and video content creators [70]. This work details the occupationally embedded nature of the platform's managerial control and corresponding workers experiences. Building on this foundation, we operationalize occupational diversity as the patterned variation in worker–technology interactions that arises from workers' embeddedness in distinct occupational contexts, where differences in tasks, roles, industry norms, and professional routines shape how technologies are materialized in workers' experiences.

However, CSCW and HCI have not yet integrated these insights into a fully occupation-centered account of platform labor, and occupation remains underexamined as a key driver of variation in workers' technology experiences. First, despite classic CSCW's focus on occupation-centered practices, recent platform-work research frequently aggregates workers into broad categories, with limited engagement with occupational embedding [56,101]. This makes it difficult to explain why identical sociotechnical systems can produce different risks and opportunities across professions, and it impedes the accumulation of occupation-sensitive design knowledge. Second, even studies that foreground occupation often examine a single occupation in isolation, obscuring which effects are occupation-specific versus platform-general, limiting comparative analysis, and constraining the articulation of design requirements [22,43]. Third, the existing literature is largely cross-sectional, leaving little multi-year evidence on how occupational conditions shape divergent experiences and outcomes in platform-mediated work over time [53].

We address these gaps by foregrounding occupation as a core conceptual and analytic framework and by employing a longitudinal, comparative research design that traces how experiences evolve over time across occupations within a single platform environment. Specifically, we focus on online freelancing on a multi-occupational platform, which affords strong analytic leverage because workers from diverse professions engage with a shared platform infrastructure, interfaces, and rules. Online freelancing is distinct from other forms of platform work in its relatively high discretion over when, where, and how work is performed and in its longer-term, project-based, knowledge-intensive engagements—conditions that render different occupational identities, skills, and norms central to everyday work practice [46,101]. By contrast, transportation and delivery platforms are more tightly governed by spatiotemporal boundaries, and crowdwork platforms are often organized around standardized microtasks, which can diminish the explanatory centrality of occupation [23]. Finally, we employ a longitudinal design to trace how these occupation-specific dynamics accumulate and change over time, addressing the field's limited multi-year evidence on occupationally differential platform work trajectories.

In this paper, we report findings from a longitudinal study of 108 U.S.-based online freelancers on Upwork. We make three contributions to CSCW. First, we empirically demonstrate that occupational contexts shape platform work experiences across four distinct dimensions: (1) online self-presentation, (2) temporal and task flexibility, (3) skill building pathways, and (4) long-term careers on digital labor platforms. Second, building on these findings, we introduce the concept of platformic occupational stratification, which captures how occupational hierarchies shape workers' uneven capacities to exercise agency under platforms' managerial control. Third, we propose design implications and future research directions for CSCW. Collectively, these contributions advance occupation-sensitive research by revealing previously understudied challenges and opportunities embedded in workers' occupational contexts. In doing so, we respond to the heterogeneous needs of professional groups navigating platform-mediated work and underscore the importance of incorporating occupational specificity into the study and design of sociotechnical systems.

The remainder of the paper is structured as follows. First, we review the literature on platform work and occupation. Next, we describe our methods, including data collection and analysis. We then present our findings, discuss their implications, and conclude the paper.

## 2   RELATED WORK

Synthesizing CSCW, organization and technology studies, and occupational sociology literature, we foreground occupation as the conceptual and analytical lens that unpacks differentiated worker–technology interactions in both conventional and platform-mediated work contexts.

### 2.1  Occupation in Platform Work

Digital labor platforms refer to privately owned online services that create a two-sided market between supply (e.g., workers) and demand (e.g., clients or companies) [31,63]. We view digital labor platforms as sociotechnical artifacts situated at the intersection of their technological functionalities and workers' social world in which interactions are mediated through their social roles such as occupations [36,55,86].

Despite frequent citation in the literature, occupational diversity remains largely nominal in studies of digital labor platforms, insufficiently developed to advance theory or guide empirical research on platform-mediated work [12,23,87]. Though few in number, some studies bring occupational dimensions of platform work to the forefront, with their increasing citation counts signaling a rising impact [43,77,81]. For example, Hoang and colleagues show that socioeconomic background drives occupation-based stratification on platforms, placing disadvantaged workers in lower-skilled roles and more affluent workers in higher-skilled ones. This, in turn, conditions workers' prospects for economic mobility [43].

However, much literature still reflects what Vallas and Schor (2020) call "theorization of platform workers as a homogeneous group" [101, p.280] or what Hoang and colleagues describe as a "uniform treatment of platform work." [43, p.638]. In the same vein, Himmelsbach and colleagues' literature review (2019) urges attention to how occupation shapes user experience, noting that scholarship often focuses instead on diversity dimensions like age, gender, and education [42]. And this gap is particularly pronounced in CSCW scholarship. A literature review (2023) reveals that less than 7% of CSCW and HCI articles (2017–2023) consider occupational contexts. Responding to this gap, Kim and Sawyer (2023) call for empirical work toward the "heterogenization of platform work", which diversifies interpretations of challenges and supports by recognizing workers' social positions, including occupation [54, p.1].

Empirical studies from CSCW-adjacent fields, such as organization and technology studies, demonstrate that occupation shapes how technologies are adopted, interpreted, and integrated into work practices [78,103]. Barley's seminal work (1986) demonstrates that identical technologies can elicit similar initial patterns of user interaction, yet these interactions often evolve in divergent ways and yield differentiated outcomes over time, contingent upon the nuanced occupational contexts in which workers' engagements with technology are embedded [9]. This underscores how occupational embeddedness shapes not only the adoption and use of technology but also the trajectories through which technologies are integrated into daily practice, as users interpret and engage with them in ways that reflect their specific occupational norms and role expectations [103]. This also means that occupation is a place for social, symbolic, and material resources that workers can draw on when negotiating boundaries or resisting external controls when interacting with technologies [11,114].

Evidence across platform studies and organization and technology studies shows that occupation matters. We now turn to traditional occupational sociology to unpack the mechanisms by which occupations structure workers' labor experiences.

## 2.2 Occupation in Traditional Work

Occupational sociologists have long argued that an individual's occupation is not just a job, but a central locus of social power that shapes life chances and class position [37,49]. Occupations are viewed as communities of practitioners defined by shared skills, knowledge, and resources, which collectively form social power [50]. Given occupation's role as a locus of social power, occupational standing is frequently used as a proxy for social class, stratifying workers' access to economic and symbolic opportunities over time [60,108,114]. Building on stratification research and professional studies [1,21,69,100], scholars advance the view that groups sharing key attributes enact "exclusionary closure", separating themselves from outsiders and monopolizing resources and opportunities [80]. This perspective translates into occupational analysis by focusing on contemporary "shared attributes" as mechanisms of social closure among occupational groups. Specific mechanisms include certifications, credentials, training, and professional memberships, each positioning individuals with differential capacities to exercise agency in labor processes [28,107]. These occupationally differential degree of capacities influence a wide range of labor aspects, including stability [47], autonomy [79], skill development [30], mobility [3], and earnings [64,88,107] among others. Accordingly, scholars argue that detailed occupations (also termed "microclasses") are the basic units of social class that often predict labor outcomes [45, p.979].

Building on these foundations, we use occupational diversity as both a concept and analytical tool. Conceptually, it foregrounds the non-uniformity of platform experiences and the occupationally embedded mechanisms that can produce stratification and inequality over time [35]. Analytically, occupational diversity allows us to approach each occupation as a comparative unit of analysis that reveals divergences structured by occupational factors rather than individual differences or platform affordances alone [9,103]. Collectively, occupational diversity renders legible the underlying commonalities and divergences across varied professional domains of platform work, providing a structured means of interpreting the otherwise heterogeneous experiences of platform-mediated labor.

## 2.3 Occupation and Platformic Management: Identity, Flexibility, Skilling, and Sustainability

Platform-mediated work simultaneously disrupts traditional occupational structures and reinforces others. On one hand, platforms are advertised as lowering the exclusionary power of professional communities by equalizing entry barriers regardless of education or class [34,41]. On the other hand, CSCW scholars often deploy the concept of platformic management to examine how digital labor platforms introduce new forms of "sociotechnical exclusionary power" [44,53]. Platformic management refers to the array of platform-driven mechanisms that automate, structure, or steer freelancers' work experiences in ways that primarily advance the platform's interests. We use platformic management as an overarching term that encompasses both algorithmic management and the co-constitutive sociotechnical mechanisms through which platforms govern labor [44, 122]. Accordingly, platform management operates through opaque search and worker-evaluation algorithms [66,74,85], discriminatory user experience and interface design choices [76,96], and platform policies and regulations that structure reputation systems [83], worker–client contractual relations, and communication arrangements [96,97]. Taken together, these strands point to a gap: we lack comparative evidence on how platform controls operate across occupations. As occupation is seen as the site where workers make sense of, resist, or accommodate direct and indirect workplace control, the effects can extend to flexibility, identity, skills, and career trajectories. We next review CSCW and related scholarship on each dimension.

*2.3.1 Platform Identity and Online Presentation.* Platform identity refers to the professional self-representation freelancers construct and maintain within platform interfaces. CSCW and related literature document that this identity is shaped and constrained by platformic management mechanisms, including reputation systems, algorithmically enforced visibility rules, and standardized profile templates embedded in platform interfaces [14,26,73,120]. In conventional settings, occupational communities with established norms and culture provide durable anchors and scripts for internal identity (self-understanding) and external presentation (signaling to others) [94,103]. On digital labor platforms, identity is pulled away from occupational communities and bound more closely to platformic management that deconstructs professional identities into quantifiable markers (e.g., client reviews, job success scores, and digital badges) [75]. These markers are then processed by platform algorithms to determine freelancers' visibility, discoverability, and ranking in client searches [52]. Thus, workers must continuously engage in identity management: curating their profiles, soliciting feedback, and strategically shaping their platformic identity while responding to the shifting mechanisms of algorithmic governance [14,19,118]. When platforms impose uniform controls, they may validate some occupational identities and flatten others, leading to divergent identity-management practices. Centering occupation may enable CSCW to examine how platform mechanisms shape identity practices across diverse occupations.

*2.3.2 Platform Flexibility.* Flexibility in digital labor platforms refers to the extent to which workers can control their working time, spatial arrangements, and autonomy in work process decision-making. Platforms market flexibility as a main benefit including control over schedules, task selection, and work environment [4]. Yet CSCW and platform studies show that platformic logics reconfigure and curtail flexibility through algorithmic and client surveillance controls [18,67,92,93]. For example, platforms reward rapid responsiveness through metrics such as Fiverr's response time [58] and Upwork's Job Success Score [57] that monitor, calculate, and showcase freelancers' response times on their profiles or in performance scores, pressuring freelancers to remain constantly available [58]. Similarly, responsiveness metrics, such as Upwork's Job Success Score, influence search rankings by prioritizing freelancers with high responsiveness to client inquiries [57]. Flexibility further varies occupationally on digital labor platforms. For instance, ride-hail drivers have low task flexibility but some spatial autonomy, while knowledge-intensive freelancers often have greater task choice and location flexibility [27,48]. Building on the evidence, we probe variations in flexibility experiences among knowledge-intensive freelancing occupations, offering more nuanced accounts to inform CSCW designs attuned to occupation-mediated flexibility.

### 2.3.3 Platform Capital and Skill.
Skill refers to knowledge, competencies, and expertise that enhance a worker's labor market value [82]. In traditional labor markets, skill is often seen as human capital acquired through formal education and accumulated professional experience, with economic rewards closely tied to credentialed achievements [72]. On digital labor platforms, however, the weight of human capital is being reconsidered: formal credentials matter less for employability and wages, while platform-specific capitals (e.g., client reviews and prior job experience) serve as key currency that influence employability [41]. Moreover, workers increasingly acquire skills via self-study, learning-by-doing, and trial-and-error, with limited institutional scaffolding [54,105,119]. This shift creates a paradox: rising pressure to upskill amid little support [54]. Crucially, this paradox is not experienced uniformly. As recent literature suggests, skill-building trajectories are profoundly shaped by workers' social and professional contexts [30,56]. This heterogenization of skill building underscores the need to examine how factors such as occupational status and other life circumstances influence access to learning opportunities and perceptions of skill demands [54].

### 2.3.4 Platform Career Sustainability.
Platform career sustainability refers to the extent to which work on digital labor platforms can offer objective prospects (e.g., career advancement, income stability) or subjective prospects (e.g., perceived satisfaction) [13,56,71]. As platform work increasingly evolves into a long-term occupational pathway rather than a temporary form of employment, scholars in CSCW and related fields have turned greater attention to long-term sustainability in platform work [13,56]. In conventional employment, occupational status predicts career sustainability: high-skill professions offer long-term contracts, protections, and career ladders, while lower-skill occupations face instability and short-term arrangements [17,29]. Yet, with the rise of digital labor platforms since the late 1990s, even well-established professions have increasingly sought work through digital labor platforms [20], giving rise to phenomena such as the "Uberization of lawyers" [116]. Unlike conventional careers buttressed by high-status occupations, platform careers demand continual navigation of income stability, reputation, and algorithmic visibility [52]. Thus, occupations once viewed as "stable" may fall under platformic control and rising precarity [47]. Precarious platform careers are increasingly widespread across occupations, yet the degree to which each occupation experiences such precarity remains uncertain. In response to the limited empirical evidence, we examine how occupational position shapes career sustainability [24,55,65].

## 3 METHODS

We present the research design of our multimethod, longitudinal, comparative study. This includes an overview of our data collection procedures, detailing the rationale behind our sampling strategy and research site selection, followed by a description of our analytical approach to both qualitative and quantitative datasets.

### 3.1 Data Collection

Our empirical foundation is based on an ongoing panel study of 108 U.S.-based online freelancers on Upwork.com. We conducted five annual data collection rounds between 2019 and 2024, spaced at approximately 12-month intervals. This longitudinal design enabled us to trace evolving patterns in the work experiences of the same 108 individuals throughout the study period.

As one of the largest and most widely used digital labor platforms [51], Upwork hosts a diverse user base spanning over 10,000 skills across more than ten broad occupational categories. We deliberately selected Upwork as our research site to facilitate comparative analysis of occupationally mediated experiences within a shared platformic environment. Unlike platforms designed for a single profession or narrowly defined sector, Upwork offers a shared infrastructural and platformic managerial setting through which workers from varied occupational backgrounds interact. This configuration enables us to analytically control for platform-level variation and focus on how differences in occupational status, norms, and role expectations give rise to divergent user experiences.

We employed purposive sampling to form our research participants. As past studies documented [91,111], probabilistic sampling is not feasible due to the limited provision of the entire user population data by platforms. We applied two inclusion criteria: a minimum of $1,000 in earnings to ensure participants' active status and experience, and U.S. location to account for cross-national variations in labor law and regulation. From this, we developed a means to randomly identify possible participants. We used a two-stage randomization process within a filtered pool of eligible Upwork freelancers. A random number generator selected both the results page and one freelancer from that page. This procedure gave each eligible freelancer an approximately equal chance of being chosen, independent of Upwork's ranking algorithms or our own discretion. Following approved protocols, we contacted each worker and asked if they would like to be hired for a one-hour job. All participants received compensation for their involvement, irrespective of whether their participation was partial or complete. In addition, we provided five-star ratings and strong positive reviews to acknowledge and appreciate their contributions.

Since Upwork periodically revises occupational categories by modifying labels and reclassifying skilled professionals, we invited a diverse sample across multiple fields to ensure broad occupational representation. With responses from all 108 workers, we analyzed job titles, project histories, skills, and qualifications from their Upwork profiles to inductively derive occupational groupings. This approach produced five overarching categories, which serve as the basis for our research findings. To ensure consistency with standardized labor classifications, we borrowed and adapted labels in accordance with the U.S. Bureau of Labor Statistics' 2018 Standard Occupational Classification System [102].

To comprehensively investigate occupational differences, we collected three complementary data sources during each wave of data collection: semi-structured interviews, survey responses, and archival platform profile data. Each participant completed a 45-minute interview focused on both professional and personal developments, including family responsibilities, employment status, and occupational context. To supplement the interviews, participants also completed a 15-minute survey capturing additional demographic information and freelancing history. Finally, we conducted a systematic review of each participant's publicly accessible Upwork profile, which included job success metrics, earned badges, skills and qualifications, project records, and client feedback.

Following initial recruitment on Upwork, we maintained contact with participants through the platform over five years to reinvite them to each wave of data collection. As is common in longitudinal studies, participants engaged with varying frequency: 21 workers completed an interview once, 25 completed two interviews, 21 completed three interviews, 13 completed four interviews, and 21 completed all five interviews. Similarly, 24 workers responded to a survey once, 20 responded twice, 21 responded three times, 18 responded four times, and 25 responded across all five waves. Altogether, this yielded 291 interviews and 327 surveys analyzed in the study.

Table 1. Participants Demographics

| Category | | Total 108 | |
| --- | --- | --- | --- |
| | | Number | Percentage |
| Gender | Female | 63 | 58% |
| | Male | 42 | 39% |
| | Unknown | 3 | 3% |
| Race | Asian | 9 | 8% |
| | Black/African American | 22 | 20% |
| | White | 56 | 52% |
| | Multi-racial | 13 | 12% |
| | Other Race | 6 | 6% |
| | Unknown | 2 | 2% |
| Education | Post-graduate degree | 38 | 35% |
| | Bachelor's degree | 46 | 43% |
| | Associates | 7 | 6% |
| | No college degree | 15 | 14% |
| | Unknown | 7 | 6% |
| Age | 18-29 | 23 | 21% |
| | 30-49 | 70 | 65% |
| | 50-64 | 13 | 12% |
| | 65 and older | 1 | 1% |
| | Unknown | 1 | 1% |
| Participant distribution by occupation | Admin Support | 27 | 25% |
| | Designers | 27 | 25% |
| | Business Analysts | 25 | 23% |
| | Information Technology | 10 | 9% |
| | Writers and Translators | 19 | 18% |
| Hourly wage by occupation (median) | Admin Support | | $20 |
| | Designers | | $45 |
| | Business Analysts | | $40 |
| | Information Technology | | $50 |
| | Writers and Translators | | $30 |

## 3.2   Data Analysis

Given the longitudinal, multi-source nature of our dataset, we foregrounded qualitative analysis and used survey and profile data as companion evidence rather than a standalone basis for inference. We first conducted thematic analysis of interviews to identify core dimensions of occupationally differentiated experiences over time. In parallel, we conducted descriptive statistical analyses of survey and profile data across occupations and rounds to assess convergence with, or divergence from, these qualitative themes.

*3.2.1 Quantitative Analysis.* The survey data analysis was based on data collected in Rounds 2–5 between 2020 and 2024. We performed a descriptive analysis of our survey and profile data, which included both numerical and categorical variables. For numerical variables, we calculated key descriptive statistics, including mean, median, standard deviation, and range. Categorical variables were analyzed using frequency and percentage distributions. This analysis aimed to identify patterns and variations, with a particular focus on differences among occupational groups within our research participants. The survey highlighted broad trends within our analytical focus, but descriptive statistics alone could not capture the mechanisms driving these differences. To deepen the analysis, we paired the survey results with interview data, using thick description and narratives to contextualize and unpack the occupationally embedded dynamics underlying the quantitative patterns.

We note that the number of respondents varies across survey rounds and occupational groups. Such variation often occurs in longitudinal research and reflects three factors. First, as discussed above, participant retention varied over the course of the study. Second, participation varied by instrument: some freelancers completed both interviews and surveys, while others provided only one type of data. Third, within the survey responses, some participants opted not to answer certain items or provided incomplete or invalid data, which were excluded from the analysis. For these reasons, the sample size for each table and indicator fluctuates across rounds. To ensure transparency, we report the number of valid responses (n) for each occupational group and survey round.

*3.2.2 Qualitative Analysis.* The interview data analysis was based on data collected in Rounds 1–5 between 2019 and 2024. To analyze the interview data, we applied a combination of grounded theory and thematic analysis [16,32]. Considering the diverse professional backgrounds of participants and the longitudinal nature of the data, we followed structured guidelines for thematic analysis [110]. The analysis was conducted systematically through the four steps.

First, familiarization and initial coding: two researchers independently reviewed transcripts from each occupational group, identifying recurring patterns and generating preliminary codes. Using NVivo 14, this phase produced 52 initial codes, which were iteratively organized into 7 parent codes and 21 child codes. For example, an early code such as creating multiple profiles evolved into the broader parent code Online Presentation, with child codes distinguishing multi-profiling within Upwork and external platform presence. Similarly, initial codes like taking on lower-skill jobs and repetitive tasks were grouped under the parent code Skilling, with child codes capturing deskilling pressures and external learning strategies. Second, code comparison and refinement: the two researchers then coded the full dataset separately and compared their work to resolve discrepancies. Through discussion, overlapping codes were merged, redundant codes removed, and ambiguous codes clarified. For instance, the initial codes work–life boundary blurring and client interruptions at night were refined into a single child code, temporal boundary management, under the parent theme Platform Flexibility. This process yielded a unified codebook that balanced breadth with analytical precision. Third, theme identification and review: the refined codes were consolidated into broader themes that captured cross-cutting dynamics across occupations. These themes were reviewed collectively in team meetings to ensure coherence across the dataset and alignment with the longitudinal design. Fourth, defining and interpreting themes: final themes were named, refined, and interpreted in light of existing scholarship. Supporting participant quotes were integrated to illustrate each theme, emphasizing how occupational context shaped divergent experiences.

## 4   FINDINGS

Our analysis resulted in four saturated themes that show distinct variations in workers' platform work experiences.

### 4.1   Occupational Diversity in Online Presentation

Findings reveal that freelancers strategically construct their online personas by leveraging platform affordances and the broader platform infrastructure to optimize self-presentation. We refer to this practice as multi-profiling: workers' strategic maintenance of multiple profiles across digital labor platforms. This approach enables freelancers to navigate platform constraints, tailor their visibility to different client expectations, and strengthen their market positioning within digital labor ecosystems.

Our analysis identifies two distinct forms of multi-profiling. First, within-platform multi-profiling occurs on a single platform (e.g., Upwork), where freelancers create more than one profile to present multiple professional personas tailored to different competencies and client demands. Second, cross-platform multi-profiling occurs when freelancers extend their presence beyond a single platform by establishing profiles on external sites (e.g., LinkedIn, personal websites, or industry-specific networks) to enhance their reputation and broaden their professional reach.

As presented in Table 2, multi-profiling practices vary across occupations. Designers engage in the highest level of within-platform multi-profiling, maintaining an average of 2.3 profiles. Conversely, Writers and Translators exhibit the most extensive cross-platform multi-profiling, averaging 6.2 external accounts beyond Upwork. In contrast, IT professionals exhibit the lowest cross-platform multi-profiling tendencies, maintaining the fewest profiles both on Upwork (1.3 profiles) and external platforms (0.5 accounts).

Table 2. Profile Count by Occupation: Upwork and External Platforms (Average across Rounds 2–5)

| Occupation | Upwork (Average profile count) | External Platforms (Average account count) | Total profiles |
|---|---|---|---|
| Admin Support (n=21) | 1.8 | 0.9 | 2.1 |
| Designers (n=22) | 2.3 | 0.8 | 3.1 |
| Business Analysts (n=18) | 1.8 | 0.6 | 2.4 |
| Information Technology (n=7) | 1.3 | 0.5 | 1.8 |
| Writers and Translators (n=23) | 2.0 | 6.2 | 8.2 |

Our interview analysis provides an in-depth account of how multi-profiling patterns differ across occupations. A key finding is that not all freelancers actively engage in multi-profiling; those in occupations with highly adaptable skill sets are more likely to establish multiple profiles, whereas those with specialized expertise often maintain a single profile to emphasize their niche competencies. These multi-profiling practices reflect distinct occupational approaches to self-presentation, influenced by the need to balance breadth and depth of expertise while aligning with client expectations on Upwork and other professional platforms.

*4.1.1 Designers: Expanding Skill Sets Through Multi-Profiling.* Designers actively engage in multi-profiling to diversify their skillsets and showcase a broad range of competencies. Rather than presenting themselves as specialists in a single domain, they strategically create multiple profiles to highlight different aspects of their expertise. P78 (Designer) offers a representative trajectory of how designers' online presentation evolves over time. When they first joined our study in 2020, they were among the earliest participants to describe leveraging a multi-profiling approach, reporting that *"I used the chance that Upwork has to create two or three specialized profiles besides your member file. It gives you the opportunity to highlight one of your niche skills, or not the main skill."* By 2022, they described how this approach translated into more offers and a broadened service portfolio, detailing how they created additional profiles to maximize what they could present to clients: *"I decided to create two more profiles that I can show to my clients. One specializes in video editing. The other specializes in email marketing. I can provide almost a very big picture of what I can do for my clients."* By leveraging Upwork's affordance of multiple profiles, designers construct distinct professional personas, each catering to specific client needs while collectively forming a comprehensive representation of their capabilities.

*4.1.2 IT Professionals: Signaling Specialized Expertise.* In contrast to designers, IT workers tend to present themselves as domain specialists rather than generalists. Their profiles foreground narrowly defined technical competencies, often signaling mastery through specific programming languages and tools, in ways that align with client expectations for IT talent. This positioning is further reinforced by industry norms that reward specialization.

P61 (Web Developer) exemplifies this pattern among nine other IT workers who constructed coherent online profiles by selectively curating the job histories they chose to display. In 2019, when we first spoke with them, they emphasized maintaining a tightly scoped profile centered on technological specialization to signal fit with what clients seek: *"There are invites coming in from clients, due to the work history, due to how well I maintain my profile. They will invite me. I will see their requirements first, if it's within the scope of my capacity or my technological ability. If they're out of scope, I would not do that."* In 2020, when we re-interviewed them about changes in their online presentation, they noted awareness of the platform's multi-profile feature but emphasized that maintaining a specialization-first profile remained essential to their continued success on the platform, reflecting broader industry norms that reward technical depth and domain expertise: *"If you don't have a specialized knowledge of the industry [...] you may not survive on this platform. That's what most of the clients are paying high wages for."*

Taken together, these findings show that freelancers' presentation work evolves over time as they adapt to platform affordances through occupation-specific strategies. Designers increasingly pursue horizontal online presentation, broadening their positioning by distributing a wider range of skills across multiple profiles to signal versatility and expand the services they can offer. In contrast, IT professionals tend to sustain vertical online presentation, reinforcing credibility by deepening signals of expertise within a narrower technical domain to match specialization-driven client demand.

## 4.2 Occupational Diversity in Platform Flexibility

Table 3. Occupational Differences in the Perceived Importance of Flexibility

| Survey Question | Occupation | Round 3 | Round 4 | Round 5 |
|---|---|---|---|---|
| Please tell us what matters to you as you pursue freelance work online: Flexibility | Admin Support | 100% (n=12) | 92% (n=11) | 90% (n=9) |
| | Designers | 93% (n=13) | 100% (n=14) | 75% (n=9) |
| | Business Analysts | 79% (n=11) | 89% (n=8) | 88% (n=7) |
| | Information Technology | 88% (n=7) | 100% (n=5) | 88% (n=7) |
| | Writers and Translators | 94% (n=17) | 100% (n=13) | 100% (n=15) |

The survey findings demonstrate that flexibility remains a central concern for freelancers across various occupations, with nearly all groups consistently reporting an importance rating of 80% or higher across the three rounds (Round 3– Round 5).

Interviews offer a more nuanced perspective, revealing the diverse ways in which freelancers experience and manage flexibility in their work on digital labor platforms. The extent of flexibility experienced in different occupations is shaped by the nature of the work and the autonomy afforded by the specific industry's expectations regarding working hours.

*4.2.1 Writers and Translators: Task Fragmentation and Work-Life Boundary Challenges.* Due to the fragmented nature of writing and translation work, writers and translators often struggle to demarcate work–life boundaries. For them, a few sentences or a paragraph may constitute the minimum unit of work they can charge for on the platform. This decomposable nature of tasks allows clients to submit "quick jobs," but it also makes freelancers' predetermined boundaries between work and non-work more permeable.

P11, a writer and translator, illustrates how sustained "always-on" expectations can make freelancing feel less viable over time. When we first spoke with them in 2020, they described repeated temporal boundary violations tied to small, ad hoc requests: *"I have to be available day and night. Sometimes people will send me a quick job at 10 o'clock at night and I'll just do it."* Two years later, they questioned the longevity of this arrangement, noting, *"I still think freelancing is not a long-term thing."* By 2025, their public platform profile shows over 250 jobs completed in total, but only two completed projects in 2024–2025, pointing to a sharp slowdown compared to earlier years.

*4.2.2 IT and Design Freelancers: Temporal and Task Flexibility.* On the other hand, data show a contrasting experience of flexibility among IT and creative freelancers over the same study period. Freelancers in these occupational groups often described experiences of working in nontraditional organizational cultures, particularly in startups or small businesses, where flexibility was a valued norm shaping schedules, communication practices, and work arrangements. P1, a data analyst, illustrates this pattern: across four interviews (2019–2022), they repeatedly emphasized how culture and hours supported their flexibility: *"That's why it's so enjoyable—you have the flexibility, you have the hours and the culture most of the time. With what I was getting before, it was like they were not big companies, but they were mostly start-ups, and the culture's pretty nice."*

Designers reported similar trajectories, frequently attributing strong temporal flexibility to clients' relaxed expectations around timelines (P5, P6, P8, P9, P30, P35, P73). For example, P35, a 3D designer, described stable schedule autonomy across the study period: in 2020, *"My work schedule is always super flexible and fluid, which is why I really love the freelance ability"*, and in 2023, *"I'm just super flexible in my schedule, and my clients are super flexible."* Designers also highlighted task flexibility, the ability to shift across project types or niches as interests and opportunities evolve. P8, a website designer explained: *"Mostly just the freedom to decide my own schedule and change my mind if I become interested in something that's more specific than website design, or a certain niche that I hadn't known about before."*

## 4.3 Occupational Diversity in Skilling

Table 4. Skilling Changes Across Occupations (Average across Rounds 4–5)

| Survey question | Occupation | Jobs requiring higher-level skills | Jobs matching my current skill level | Jobs requiring lower-level skills |
|---|---|---|---|---|
| Over the last 12 months, what percentage of your Upwork jobs fall into each of the following categories? | Admin Support (n=11) | 9% | 71% | 20% |
| | Designers (n=12) | 17% | 58% | 25% |
| | Business Analysts (n=11) | 14% | 39% | 48% |
| | Information Technology (n=7) | 21% | 53% | 26% |
| | Writers and Translators (n=11) | 8% | 55% | 37% |

We examine how well online freelancers advance their skills through a lens of skill changes. By analyzing the distribution of jobs requiring higher, matching, or lower skill levels across different occupations, we aim to identify trends in upskilling, stability, and deskilling. This analysis offers insights into how different freelance professions experience skill evolution, whether they trend toward upskilling, skill maintenance, or deskilling over time.

*4.3.1 Most Freelancers Perform Work at Their Current Skill Level.* As shown in Table 4, across all occupational groups, the largest share of freelance jobs falls into the "Jobs matching my current skill level" category. Despite some evidence of upskilling or deskilling, freelancers in four of the five occupational groups (administrative support, design, information technology, and writing and translation) reported taking jobs aligned with their existing skill sets during the past 12 months, suggesting an overall tendency to remain within their established areas of competence.

Among all occupational groups, Business Analysts reported the highest percentage (48%) of jobs requiring lower-level skills. Information Technology freelancers reported the highest proportion (21%) of jobs requiring higher-level skills. Administrative Support freelancers exhibited the highest level of skill stability, with 71% of their jobs matching their current skill level, the largest percentage across all groups.

Table 5. Occupational Differences in Formal Schooling Value (Average across Rounds 4–5)

| Survey question | Occupation | Not at all valuable | Somewhat valuable | Valuable | Very valuable |
|---|---|---|---|---|---|
| How valuable to your freelance work is your formal schooling? | Admin Support (n=24) | 25% | 38% | 17% | 20% |
| | Designers (n=24) | 8% | 42% | 21% | 29% |
| | Business Analysts (n=14) | 21% | 43% | 21% | 15% |

| | | | | |
|---|---|---|---|---|
| Information Technology (n=10) | 0% | 56% | 44% | 0% |
| Writers and Translators (n=25) | 8% | 25% | 25% | 42% |

*4.3.2 Formal Education: Moderately Useful Overall, Highly Valued by Writers, but Less Crucial for IT Freelancers.* As presented in Table 5, across all occupational groups, "Somewhat valuable" was the most common response, indicating that formal education provides a foundational skillset but is often supplemented with alternative learning pathways. While relatively few respondents considered formal schooling "Very valuable", there was also a notable group in each occupation who rated it "Not at all valuable".

Among all occupational groups, Writers and Translators had the highest percentage of respondents (42%) rating formal schooling as "Very valuable", followed by 25% stating it was "Valuable". This suggests that academic training plays a more direct role in freelance careers in writing and translation compared to other fields. The relatively low percentage of respondents (8%) stating that formal education was "Not at all valuable" further supports this conclusion. A particularly striking finding emerged from freelancers in Information Technology roles. None of the respondents rated formal education as "Very valuable", and an overwhelming majority (56%) considered it "Somewhat valuable", with the remaining 44% stating it was "Valuable".

*4.3.3 Deskilling in Writing and Translation and Continuous Upskilling in IT.* Interviews reveal the occupation-specific nature of skilling experiences among freelancers. While survey data indicate that a larger proportion of writers and translators report engaging in jobs requiring lower-level skills and place considerable value on formal education, the interview findings provide a deeper understanding of this trend. Interviews reveal that freelancers in this field face challenges due to the repetitive nature of their tasks and the limited opportunities for skill development through the work they perform on digital labor platforms. The findings suggest that the prevalence of low-level skill jobs may be influenced by the specific characteristics of writing and translation work available on these platforms. Additionally, the emphasis on formal education stems from the need to rely on academic training, as the work itself does not provide sufficient opportunities for skill development in this occupation.

Freelancers pointed out the monotony of many tasks, noting that a significant portion of their work could be automated. P47, a translator who has freelanced on Upwork since 2017, expressed this concern in 2022: "There's lots of repetitive stuff in translation, so what I do try to do is, any time I run across something boring and repetitive, find some way to automate it and speed it up." They completed over 200 jobs between 2017 and 2025. However, about 93 percent of those jobs were secured before 2022, and since 2022 their workload has fallen to fewer than five jobs per year, which substantially reduces their opportunities to skill through work.

Additionally, freelancers often find that clients prioritize speed and cost over quality. One translator commented, "We live in a day and age where language is not as valued as it once was. What that reflects to me is that speed is priority, not quality." (P39, Translator). Another freelancer noted, "There are many, many employers out there who are looking to get the cheapest possible labor, which makes sense, but it's at a very low cost." (P41, Writer). This focus on inexpensive and quick labor, combined with the repetitive tasks, presents a substantial barrier to skill development.

Reflecting further on the lack of skill-building opportunities within their work, several freelancers reported not experiencing upskilling through their day-to-day tasks (P23, P13, P47). As a result, many freelancers seek learning through external resources. One freelancer shared, *"I* *try to seek out teachers and since I'm no longer in college, that is one way I can do that is looking for online courses, for teachers that are teaching in that regard."* (P29, Writer).

In contrast, IT freelancers, who reported the highest percentage of jobs requiring higher-level skills and placed less emphasis on formal education in the survey, offer a more detailed perspective on this trend in the interviews. This group reports having opportunities for skill development through the tasks they perform on digital labor platforms (P1, P46, P61, P75, P86, P87, P90). These findings suggest that they are able to secure and engage in jobs that demand advanced skills, thus relying more on the knowledge and expertise acquired through freelance work rather than on formal education.

Unlike traditional jobs, where employees are often limited to specific tasks and may not interact with end clients, IT freelancers on platforms have direct access to clients. This interaction provides them with opportunities to stay updated on industry trends and emerging technologies, helping them develop skills for professional advancement. P61, a web developer who has freelanced since 2018, offers a representative account: *"The opportunity to work on a variety of projects gives you an opportunity to advance in your skills. Whereas, on the contrary, in companies many times you are bound to do a specific job. On Upwork, working for the end clients, you are valued in the eyes of your clients, whereas in a regular job, you may not deal with your end clients. Your managers, senior managers, delivery managers, they will be dealing with the end clients. For example, your consumer, the website utilizers, software utilizers, you get the chance to know them, how the industry will be, leading towards what will be the future, what technology will be coming to market."* From 2018 to 2025, P61 completed more than 100 jobs. Unlike the translator and writer trajectories discussed earlier, their job-taking pattern remained broadly stable across this period.

## 4.4  Occupational Diversity in Platform Work Sustainability

We report cross-round changes related to platform work sustainability. This section relies on survey data because the three measures used to understand sustainability were collected consistently across rounds, allowing for longitudinal comparison across occupational groups in ways that the qualitative findings were not intended to support. Rather than offering definitive conclusions, the findings presented here illuminate broader patterns that help contextualize participants' occupational experiences within the larger question of platform work sustainability.

We examine platform work sustainability through three indicators: (1) Earnings stability over time, measured by longitudinal changes in monthly income, reflecting how financial stability fluctuates across occupations; (2) Reliance on Upwork income over time, assessed through temporal shifts in the proportion of freelancers' monthly earnings derived from the platform, indicating variations in financial dependence across different occupations; and (3) Perceptions of Upwork as a long-term career, evaluated through changes in freelancers' long-term work plans, revealing evolving expectations and whether platform work is increasingly viewed as a sustainable or temporary career path.

The three findings highlight distinct occupational trends: Overall, most occupations experienced a downward trend in platform earnings, dependence on platform income, and the perception of platform work as a long-term career. IT professionals, despite initially having the highest earnings, experienced the most significant income decline over time. Admin Support workers, despite earning less than most other occupational groups, became increasingly dependent on Upwork as an income source. And Writers and Translators exhibited one of the most pronounced declines in their perception of Upwork as a viable long-term career.

4.4.1 Platform Earnings.

Table 6. Longitudinal Shifts in Monthly Earning

| Survey Question | Occupation | Round 2 | Round 3 | Round 4 | Round 5 |
|---|---|---|---|---|---|
| In a typical month, approximately how much do you earn via Upwork? | Admin Support | $800 (n=10) | $500 (n=11) | $450 (n=11) | $350 (n=12) |
| | Designers | $350 (n=12) | $500 (n=10) | $50 (n=12) | $300 (n=14) |
| | Business Analysts | $775 (n=8) | $450 (n=6) | $75 (n=8) | $100 (n=8) |
| | Information Technology | $1,500 (n=7) | $500 (n=4) | $500 (n=6) | $125 (n=5) |
| | Writers and Translators | $400 (n=15) | $225 (n=21) | $250 (n=15) | $275 (n=14) |

Table 6 illustrates the longitudinal changes in median monthly earnings across five occupational categories, based on responses collected over four survey rounds (Round 2 to Round 5). This question investigates changes in earnings over time to assess the financial stability/insecurity experienced by freelancers in different occupational fields. Analysis reveals three trends. Overall, earnings exhibited a downward trend across all occupations, with notable fluctuations in certain fields. Among them, IT professionals, who reported the highest initial monthly earnings of $1,500 in Round 2, experienced the most substantial decline, dropping to $125 by Round 5 (-91.7%). In contrast, designers, who had the lowest reported earnings at $350 in Round 2, experienced the most modest reduction, to $300 in Round 5 (-14.3%).

4.4.2 Platform Income Dependency.

Table 7. Longitudinal Shifts in Upwork Income Dependency

| Survey question | Occupation | Round 2 | Round 3 | Round 4 | Round 5 |
|---|---|---|---|---|---|
| On average, what percentage of your monthly earnings comes from Upwork jobs? | Admin Support | 19% (n=11) | 20% (n=11) | 18% (n=12) | 25% (n=11) |
| | Designers | 19% (n=11) | 15% (n=12) | 15% (n=12) | 10% (n=11) |
| | Business Analysts | 16% (n=6) | 20% (n=9) | 5% (n=7) | 8% (n=8) |
| | Information Technology | 38% (n=4) | 10% (n=6) | 8% (n=3) | 2% (n=7) |
| | Writers and Translators | 20% (n=21) | 25% (n=16) | 20% (n=13) | 12% (n=12) |

Table 7 presents the longitudinal shifts in the proportion of freelancers' monthly earnings derived from Upwork across five occupations over four survey rounds (Round 2 to Round 5). This analysis explores the evolving reliance on Upwork as a primary income source. The findings reveal three trends. Overall, many occupations tend to experience a decline in the share of monthly earnings generated through Upwork. The most pronounced decrease was observed in Information Technology, where dependency declined from 38% in Round 2 to 2% in Round 5 (-36%). In contrast, Admin Support exhibited a gradual increase, rising from 19% in Round 2 to 25% in Round 5 (+6%).

### 4.4.3 Long-term Platform Work Plan.

Table 8. Longitudinal Shifts in Perceptions of Upwork as a Long-Term Plan

| Survey question | Occupation | Round 2 | Round 3 | Round 4 | Round 5 | Total N (R1-5) |
|---|---|---|---|---|---|---|
| I see Upwork as part of my long-term plan | Admin Support | 73%[1] (n=8) | 92% (n=11) | 75% (n=9) | 64% (n=7) | 11,12,12,11 |
| | Designers | 55% (n=6) | 60% (n=9) | 43% (n=6) | 25% (n=3) | 11,15,14,12 |
| | Business Analysts | 67% (n=4) | 21% (n=3) | 22% (n=2) | 25% (n=2) | 6, 14, 9, 8 |
| | Information Technology | 50% (n=2) | 75% (n=6) | 80% (n=4) | 63% (n=5) | 4,8,5,8 |
| | Writers and Translators | 57% (n=4) | 67% (n=12) | 71% (n=10) | 27% (n=4) | 7,18,14,15 |

Table 8 shows the changes in online freelancers' perceptions of Upwork as a long-term career across five occupations over four survey rounds (Round 2 to Round 5). This analysis investigates the extent to which freelancers consider Upwork a sustainable career path. The findings reveal three trends. Overall, the perception of Upwork as a viable plan declined across most occupations. Management analysts experienced one of the most significant drops, with long-term views decreasing from 67% in Round 2 to 21% in Round 3 (-46%). Likewise, writers and editors saw a sharp decline, with those considering Upwork a long-term option falling from 71% in Round 4 to 27% in Round 5 (-44%). In contrast, software and web developers remained the most stable, increasing from 50% in Round 2 to 80% in Round 4 (+30%) before a slight decline to 62% in Round 5 (-18%).

## 5    DISCUSSION

Drawing on a comparative longitudinal study, we demonstrate how occupational context shapes freelancers' experiences along four dimensions. From these findings, we develop the concept of platformic occupational stratification and its four mechanisms to capture how occupational contexts shape uneven experiences within platform-based work. We also discuss how workers from different occupational backgrounds navigate or resist platformic occupational stratification. Building on this discussion, we outline implications for future CSCW research and design practices.

### 5.1    Occupational Diversity

We advance the conceptualization of occupational diversity as patterned variation in platform-mediated work, evident in the occupation-specific ways freelancers engage with, navigate, and interpret platforms' sociotechnical arrangements to perform work and sustain careers. In doing so, we respond to the field's increasing recognition yet ongoing undertheorization of occupational diversity in CSCW research on platform labor [43,53,101].

Our analysis makes it clear that participants from distinct occupational domains demonstrate differentiated practices and outcomes across four dimensions: online presentation, flexibility, skilling, and platform sustainability. By foregrounding the mediating role of occupational context, we surface forms of differentiation that are obscured by generalized accounts of platform labor, which often neglect the occupational embeddedness of workers' practices. As an analytical construct, occupational diversity illuminates how freelancers' engagements with technology are shaped not solely by individual preferences or platform affordances, but by their position within occupational structures. Attending to this occupational embeddedness allows CSCW scholarship to move beyond reductive or universalizing models of platform labor [101], and instead account for how distinct professional communities negotiate, resist, and appropriate digital infrastructures in contextually situated ways.

---

[1]    Percentages in Table 8 represent the proportion of respondents who selected the survey item (n = number of respondents selecting the item; N = total participants in each occupational category).

## 5.2 Platformic Occupational Stratification

Connecting occupational diversity with CSCW accounts of platformic management [44] and occupational sociology's view of occupations as loci of social power and closure [49,107,108], we introduce platformic occupational stratification. Occupational diversity describes patterned, occupation-specific engagements with shared platform infrastructures. In contrast, platformic occupational stratification explains how platforms' information systems and design structures translate occupational diversity into stratified work processes and outcomes that mirror and often amplify long-standing occupational hierarchies in traditional labor markets. In conventional labor markets, occupational communities, credentials, and training structure access to stability, autonomy, and financial and professional rewards [2,47,64,79,108]. In platform-mediated work, these classic mechanisms are reworked through platforms' information and design arrangements, producing new forms of sociotechnical closure and opportunity.

Findings demonstrate that Upwork's shared infrastructure yields stratified work experiences and outcomes across four key domains: (1) online self-presentation, (2) flexibility, (3) skilling, and (4) platform career sustainability. We observed that the very same information, design arrangements and client interaction intersect with occupational diversity to produce uneven benefits and burdens. As summarized in Table 9, we unpack the mechanisms driving platformic occupational stratification across four key dimensions of platform work.

Table 9. Four Mechanisms of Platformic Occupational Stratification

| Mechanism | Definition | Implications for Platform Work |
|---|---|---|
| Identity Codifiability | Degree to which tasks, processes, value claims can be encoded in platforms' templates, data schemas, and re-expressed in profiles. | Higher codifiability expands legibility bandwidth → enables fuller activation of profile affordances → produces improved market signals and employability. |
| Temporal Negotiability | Extent to which tasks can be modularized and completed asynchronously, decoupled from clients' communication demands and oversight. | Higher negotiability → permits asynchronous progress → facilitates alignment with personal rhythms → enhances boundary control → leads to flexibility realized as actionable autonomy. |
| Upskilling Capacity | Degree to which an occupation's tasks and client mix enable expansive vs. reactive skill building. | Higher capacity → workers take on stretch projects → engage with cutting-edge tools and industry trends → experience platform work as an upskilling accelerator. |

| Platform Sustainability | Degree to which an occupation achieves identity codifiability, temporal negotiability, upskilling capacity to shape long-term career durability on or off platforms. | Signal salience (algorithmic visibility), schedule sovereignty (temporal control), and learning conversion (upskilling to higher rates) can contribute to improved career sustainability. |
|---|---|---|

*5.2.1 Identity Codifiability.* Prior studies conceptualize platformic self-presentation as a strategic, dramaturgical performance through which workers render their activities into legible, platform-mediated representations to communicate a professional self to audiences (e.g., clients) and enhance employability [14,33]. We advance past studies by demonstrating that occupation structures how and how far workers can activate platformic self-presentation possibilities, generating systematic uneven platform work experience across occupations that might persist beyond presentation itself.

Standardized profile features (e.g., fields for job titles, skills, portfolios, and client feedback) ostensibly provide a common set of action possibilities for developing and managing a professional persona, yet occupational position shapes how, and how far, workers can mobilize these configurations. Specifically, an occupation's identity codifiability, the degree to which its tasks, processes, and value claims can be encoded within platform templates and data schemas and re-expressed in profiles, shapes the legibility of workers' professional selves on digital labor platforms.

For example, high-codifiability occupations (e.g., designers, voice-over artists) produce visually demonstrable artifacts (e.g., rich visual and audio portfolios) that map neatly onto platform templates and provide relatively legible signals to prospective clients. By contrast, low-codifiability occupations (e.g., administrative support) lack easily visualized ways to display work processes or the value of outputs, leaving workers with less control over professional presentation and differentiation. Thus, occupational location shapes the extent to which ostensibly universal profile affordances can be activated and, in turn, the capacity for platformic self-presentation.

Linking these findings to research on labor-market signaling and occupation-based employability [95,108], we suggest that variation in codifiability and the concomitant differential legibility bandwidth can produce cumulative effects. Occupations with higher codifiability are more likely to experience platformic identity amplification, translating into greater platform visibility and employability, whereas lower-codifiability occupations face constrained opportunities. Over time, these dynamics can compound through reputation mechanisms (e.g., job history, client reviews), generating path-dependent advantages in earnings and professional development.

*5.2.2 Temporal Negotiability.* Existing literature shows that the platform promise of "work whenever, wherever" often collides with client deadlines, global time zones, and algorithmic cues, yielding flexibility in principle but uneven temporal autonomy in practice [18,67,68,93]. We reveal the hidden narratives of platform-afforded flexibility by showing that although all freelancers nominally have control over when and where they work, the degree of temporal flexibility and its implications vary by occupation.

While the digital labor platform provides a shared infrastructure for flexibility, industry norms, task characteristics, and client expectations in each occupational domain shape how flexibility is enacted and experienced. Specifically, we highlight occupational temporal negotiability: the extent to which tasks can be modularized and completed asynchronously, decoupled from clients' real-time demands. Temporal negotiability can be shaped by three factors: task modularity, synchrony requirements, and client oversight norms.

Occupations with high temporal negotiability (e.g., technical and creative freelancing) are organized around project-based, relatively nondecomposable tasks that permit asynchronous progress and allow workers to align schedules with personal rhythms. By contrast, occupations with constrained temporal negotiability (e.g., administrative support, writing/translation) more often confront boundary permeability and off-hours demands, driven by small, decomposable tasks and quick-turn requests. This task modularity encourages clients to treat labor as on-demand, which in turn creates synchrony expectations and strengthens client oversight norms (e.g., granular deadlines, frequent check-ins, time tracking). Consequently, even with identical platform tools, the unit of work (e.g., a paragraph, a page) invites "just-in-time" requests and continuous-availability expectations, collectively compressing workers' schedule control. Taken together, the same scheduling affordances yield occupationally stratified flexibility: roles with higher temporal negotiability convert nominal platform freedom into actionable autonomy, whereas roles with lower temporal negotiability experience flexibility primarily as responsive availability with constrained schedule control.

*5.2.3. Skilling Capacity.* Traditional accounts treat skill as human capital, knowledge and competencies accumulated through formal education and experience, with returns tied to credentials [72,82]. On digital labor platforms, however, the currency of advancement shifts toward platform-specific capital, datafied signals such as prior jobs and client reviews, while most skill building occurs outside formal education and amid thin institutional scaffolding [41,55].

We extend this literature by showing that platform-mediated skill formation does not constitute an entirely new logic; it also reproduces occupational stratification long observed in conventional labor markets. We term this pattern occupationally stratified upskilling capacity: the extent to which an occupation's task ecology and client mix enable expansive (frontier-reaching) versus reactive (maintenance-oriented) skill development on platforms. Capacity can be shaped by three factors: the novelty gradient of typical projects (exposure to new tools/domains), project scope (opportunities to integrate new capabilities rather than execute narrow fragments), and the conversion of newly acquired skills into opportunities via platform capital.

Digital labor platforms operate as both accelerators and constraints. In expansive upskilling (accelerator model), common in high-tech and emerging fields (e.g., data science, IT), workers strategically pursue upskilling through stretch projects involving cutting-edge tools and novel problem spaces. In this case, digital labor platforms function as a site of real-time learning and rapid portfolio iteration, with completions quickly translating into credibility and access to more complex work. In reactive upskilling (constraint model), typical of established or lower-status fields (e.g., translation, administrative support), workers confront commoditization from automation and intensified price competition, often without commensurate gains in platform capital, pay, or status. In effect, platforms shape divergent occupational skilling trajectories, enabling advancement in high-demand domains while channeling more defensive forms of skilling in routine or traditional fields.

*5.2.4. Platform Work Sustainability.* Past studies on platform-mediated careers document both the promise and burden of sustaining long-term livelihoods online [13,56]. Classic labor research links career sustainability to occupational status and institutional protections, yet platformization of work and careers has unsettled even high-status professions [10,116]. What remains under-specified is how occupational differences translate platform affordances into durable (or fragile) career trajectories. We advance platform-mediated careers literature by showing that platform sustainability is occupationally stratified through the combined action of three mechanisms established in our earlier sub-sections and findings.

For instance, IT and design roles align with high codifiability (artifact-anchored portfolios), high temporal negotiability (longer project-based, asynchronous work), and expansive upskilling (stretch projects in novel learning experiences). This triad may enable such occupational groups to maximize algorithmic visibility and credible signaling, support long-term contracts, and convert learning into rate growth, eventually more sustainable, upwardly mobile careers. By contrast, administrative support and writing/translation with low codifiability (less displayable artifacts), constrained negotiability (microtasks, quick turns), and reactive upskilling (defensive tool adoption). These conditions may depress visibility, slow reputation compounding, and produce episodic gigs and price competition and thus fragile sustainability.

Findings show an overall downward trajectory across three career sustainability indicators (monthly earnings, the share of total income derived from Upwork, perceptions of Upwork as part of one's long-term career). The slopes, however, are occupation-specific. IT respondents began with the highest on-platform earnings (2020, Round 2) but ended with the lowest platform dependence (2024, Round 5), suggesting diversification to other markets or off-platform retainers. This off-platform career mobility may be attributable to high codifiability, high temporal negotiability, and expansive upskilling. By contrast, Admin Support workers saw declining earnings alongside rising dependence on Upwork over the same period, suggesting limited scope to shift income off-platform under conditions of low codifiability, constrained negotiability, and reactive upskilling. Taken together, the data reveal the occupationally stratified nature of platform sustainability: where the triad aligns, workers convert platform affordances into mobility and durable career pipelines; where it does not, reliance intensifies even as returns decline.

## 5.3 Occupationally Situated Worker Agency

Findings also reveal that platform workers' enactment of agency, such as self-presentation and skilling, is shaped by their occupational contexts, highlighting occupationally differentiated resistance to platformic management. By applying an occupational diversity lens, we show that workers leverage platform affordances in distinct ways that reflect their professional norms, task structures, and status hierarchies. This insight advances sociotechnical perspectives in CSCW and HCI, which conceptualize algorithms and interfaces as simultaneously constraining and enabling worker action [66]. For instance, Zhang and colleagues argue that algorithmic management restricts worker autonomy, prompting resistance through system manipulation [117]. We build on this by demonstrating how such resistance is inflected by occupation: freelancers develop distinct, occupation-specific strategies to navigate platformic management.

Designers, for example, engage in within-platform multi-profiling, maintaining several Upwork profiles to display specialized competencies (e.g., video editing vs. email marketing). These profiles are finely tuned to algorithmic search functions, incorporating strategic keywords and portfolio assets aligned with client demand [75]. In contrast, writers and translators selectively foreground high-demand skills such as ghostwriting, curating their profiles to align with platform-promoted narratives of value. Admin support workers, who often lack portfolio-friendly outputs, tend to rely on metrics like job volume and ratings, usually through a single, generalist profile. These divergent strategies represent subtle forms of resistance against the platform's reductive classificatory schemes. Rather than conforming to algorithmic representations, freelancers actively curate their visibility, manipulate profile metrics, and construct personal brands that challenge the platform's narrow framing of their labor.

Occupational distinctions also emerge in skill development strategies. Freelancers in high-tech sectors (e.g., software development or data science) often engage in proactive upskilling, using platform projects as opportunities to stay up to date with industry trends and in-demand skill requirements. In this context, Upwork becomes a gateway to cutting-edge tools and evolving industry standards. By contrast, translators and administrative workers typically adopt reactive or compensatory skilling, learning adjacent tools in response to eroding demand in their core fields. This learning is often self-directed, unsupported, and uncertain in payoff. Writers, in particular, express a sense of burden in having to continuously adapt just to stay competitive, suggesting that upskilling, rather than being empowering, can become an exhausting necessity.

Together, these insights challenge overly optimistic narratives that frame platform-mediated learning as uniformly empowering [25]. Instead, we show that skilling is a contested terrain, one where resistance to deskilling is deeply shaped by occupational structures and unevenly rewarded. This demands more granular attention to how learning and visibility are differently accessible across professions, and under what conditions they translate into meaningful platform work experiences.

Revealing such differentiated forms of agentic practice sharpens our discussion of design and research directions, specifically the need to attend more carefully to occupational groups whose capacity to resist or adapt to platformic management is constrained by structural limitations embedded in their professions. These findings underscore the importance of developing occupation-sensitive design interventions and research frameworks that do not assume uniform agency across platform workers.

## 5.4 Design and Research Implications for CSCW

Our discussion of platformic occupational stratification reveals that ostensibly uniform digital infrastructures produce differentiated occupational experiences across four key domains. These findings underscore how platform management mechanisms interact with pre-existing occupational hierarchies in ways that exacerbate asymmetries in freelancers' experiences. This insight calls for a reorientation of CSCW design and research practices to engage more directly with occupational differences. We propose two interrelated trajectories to advance CSCW scholarship and practice.

*5.4.1. Occupation-Sensitive Design Interventions.* (1) Enhancing representational affordances: Occupational groups possess uneven capacities to make their expertise visible within existing platform structures. Creative professionals such as designers often benefit from visually rich portfolios and tool-based signaling. In contrast, workers in administrative or linguistic domains face limited means of conveying competence and credibility. To address this imbalance, platform design should support diverse representational affordances that align with occupationally specific norms and capabilities. This could include structured narrative templates that scaffold self-presentation for each occupational group or validated metrics of efficiency and accuracy tailored to non-visual domains. For instance, platforms could allow IT professionals to integrate external repositories such as GitHub, enabling more tailored and comprehensive demonstrations of technical expertise. Similarly, for writers, platforms might support portfolio links to published work or offer readability scores and turnaround benchmarks.

(2) Embedding occupational skilling pathways: Our data show that opportunities for skill development are unevenly distributed across occupational groups. Freelancers in technical fields such as IT often integrate learning into their rapidly evolving workflows, gaining exposure to new tools and technologies as part of their tasks. Conversely, those in occupations like writing or administrative support must pursue self-directed, unsupported learning to remain competitive. Current platform features (e.g., job postings structured exclusively around client-defined requirements) tend to obscure the developmental value of project-based work. We propose that platforms should redesign job postings to signal skilling opportunities, making explicit what freelancers can expect to learn or improve upon through project engagement. For example, listings could include information about tools to be used or skill domains that a task is likely to enhance. This shift from an extractive to a mutually developmental model could support more sustainable and inclusive skill-building, especially in occupations where formal upskilling pathways are less visible or accessible.

*5.4.2. Occupation-Sensitive Research Directions.* (1) Theorizing worker resistance as an occupationally situated practice. Findings offer initial evidence that workers are not simply passive targets of platform control; rather, they actively exercise agency to navigate platform constraints in ways that often reflect their occupational roles, resources, and priorities. Although beyond the analytical scope of this paper, one illustrative practice we observed is multi-occupational profiling, in which freelancers use multiple profiles to cross occupational boundaries and pursue resources and opportunities associated with other fields. For instance, a designer may create a secondary profile as a data analyst to broaden eligibility for different project types and enhance visibility within matching algorithms that may favor in-demand skill sets. This occupational boundary spanning suggests a deliberate tactic through which workers reconfigure professional identities and opportunity structures to work around rigid occupational classifications embedded in platforms and broader labor-market norms. Future CSCW research can build on this initial evidence by studying how occupationally situated agency emerges in practice, how workers navigate platform-enabled constraints and opportunities, and what governance or design interventions can better support workers' strategic efforts.

(2) Contextualizing occupational diversity across the broader platform-work landscape. Our analysis centers on knowledge-work occupations within a single platform, but this group represents only one segment of a much larger platform-labor landscape [113] Future work should evaluate how transferable our insights are to other forms of platform labor (e.g., ride-hailing, delivery, care work), where occupational diversity and stratification may take different shapes. Moving beyond single-platform analyses, cross-platform comparisons of occupationally differentiated experiences can sharpen occupation-sensitive CSCW theorizing by revealing when occupation best explains differences in worker experiences, when broader structural forces (e.g., platform design, governance, or regulation) dominate, and how interventions should be recalibrated across platform labor ecologies.

(3) Advancing longitudinal methods. Although CSCW scholars have emphasized the value of longitudinal work, longitudinal approaches are still not widely adopted in the community [59,106]. Our multi-wave evidence contributes to CSCW's methodological push for longitudinal research, but it also makes visible practical constraints that future work could address. One key challenge is attrition across waves: as participation declines or becomes uneven, the amount of valid longitudinal data shrinks, which can weaken analytic power and shape the contours of the findings. Recognizing attrition as a common challenge in panel-based longitudinal research, we used a mixed-methods approach to fill gaps created by uneven participation by drawing on complementary data sources (e.g., platform profiles and professional social media traces collected with consent). Even so, more stable participation would provide a stronger empirical foundation and further strengthen the validity of longitudinal claims. Future CSCW research would benefit from methodological development that makes multi-year panels more feasible and sustainable, for example, by lowering participation barriers and cultivating engagement through community-building and ongoing communication [98]. Collectively, contributions along these lines can expand the community's capacity to carry out robust longitudinal research and, in turn, better examine how technologies, platforms, and work practices evolve over time.

It is also worth noting that CSCW has a longstanding tradition of longitudinal inquiry that extends beyond panel-based or multi-wave designs like the one used in this study. This broader tradition includes ethnographic and other sustained forms of engagement, each of which reflects the field's enduring understanding that many sociotechnical phenomena unfold gradually and can be meaningfully examined over time. In this sense, the value of longitudinal CSCW research lies not in any single methodological form, but in its capacity to capture processes, relationships, and structural conditions that become visible only through sustained inquiry. We therefore hope future efforts continue to acknowledge, support, and expand the value of longitudinal CSCW research across this wider methodological range.

(4) Exploring occupational labor practices in the context of broader social and technological change. A broader contribution of this study is to foreground occupation as a key condition shaping how sociotechnical systems differentially affect workers' labor experiences. While our empirical analysis is anchored in a single sociotechnical system, we see this as a starting point for a broader line of inquiry. Future research should move beyond occupationally mediated experiences within particular sociotechnical systems to examine how wider contextual changes surrounding those systems are also mediated through occupational heterogeneity. For example, prior work on online freelancing during the COVID-19 pandemic similarly highlights occupational differences in workers' experiences, suggesting that large-scale social disruptions can reverberate unevenly across platform labor in occupation-specific ways [121].

Looking ahead, other forms of external disruption warrant similar attention. Most notably, the growing introduction of artificial intelligence (AI) raises urgent questions for CSCW scholarship at the intersection of occupation, labor, and technology. One important avenue for future research is to examine occupational heterogeneity in AI-mediated labor: how such technological changes produce uneven effects across occupational groups, what new challenges and opportunities they create, and how they may intensify, redistribute, or transform the forms of stratification identified in this paper and beyond. These questions also prompt inquiry into whether AI further deepens inequalities in skill, worker agency, professional identity, and longer-term socioeconomic mobility, or whether it may also create new possibilities for occupational adaptation and resistance. They also invite research into how sociotechnical design might better mitigate occupation-specific harms while supporting more equitable forms of adaptation in the context of AI. In this sense, the insights from our study can be understood as part of a broader research agenda concerned with investigating and responding to occupationally differentiated labor experiences under conditions of social and technological change.

## 6 CONCLUSION

In response to growing calls for an occupation-sensitive perspective on platform work, this study advances CSCW scholarship by foregrounding occupation as a key axis of variation in platform-mediated labor. Drawing on a multimethod longitudinal comparative study of 108 freelancers on Upwork, we offer three key contributions. First, we demonstrate how occupational contexts interact with platform design and technological infrastructures. Our findings reveal four distinct patterns in workers' experiences across occupations, particularly in their capacity to manage professional self-presentation, exercise temporal and task flexibility, and engage in skill development and sustain long-term careers. Second, we introduce the concept of platformic occupational stratification and four mechanisms to explain how platform management intersects with occupational hierarchies, shaping differential experiences of platform work. Third, we provide design and research implications for developing sociotechnical systems that are responsive to the situated needs and constraints of diverse occupational groups. Taken together, these contributions strengthen CSCW scholarship's capacity to account for the heterogeneity of platform labor and provide a foundation for more occupation-sensitive design practices.